# On the Impossibility of a Poincare-Invariant Vacuum State with Unit Norm

Jeremy Berkowitz[*]
University of Houston
February 6, 2008

**Abstract:** In the standard construction of Quantum Field Theory, a vacuum state is required. The vacuum is a vector in a separable, infinite-dimensional Hilbert space often referred to as Fock space. By definition the vacuum wavestate depends on nothing and must be translationally invariant. We show that any such translationally-invariant vector must have a norm that is either divergent or equal to zero. It is impossible for any state to be both everywhere translationally invariant and also have a norm of one. The axioms of QFT cannot be made internally consistent.

## 1. Introduction

The axioms of Quantum Field Theory require the existence of a unique vacuum wavestate which is a vector in an infinite-dimensional Hilbert space, often referred to as Fock space. This vacuum state is the unique state which is everywhere translationally invariant under the representation of the Poincare (inhomogeneous Lorentz) group. We show that any such translationally-invariant vector in Hilbert space must be constant in a particular sense. As a result, its Hilbert space norm must be divergent or equal to zero.

Unfortunately, the axioms of Quantum Field Theory require the vacuum state have a norm of 1. Since a finite nonzero norm is mathematically inconsistent with translational invariance, a contradiction is established. A consistent theory cannot require a translationally (and hence Poincare) invariant element with unit length. We make no use of Lorentz transformations in this paper whatsoever, because a contradiction is established by using just pure translations of the inhomogenous Lorentz group.

[*] Address correspondence to: jberkowitz@uh.edu. I gratefully acknowledge the comments of Roman Jackiw, John McGreevy and Yoichiro Nambu. Any errors or inaccuracies are solely those of the author.

As a result of this contradiction, the axioms of Quantum Field Theory as described, for example in [2, 3], cannot be formulated. To derive a theory that is internally consistent, one of the axioms must be weakened. We consider both possibilities: either the vacuum is not of unit norm, or there is no unique vacuum state. A short discussion suggests that the only workable choice is to retain the assumption that all states are unit rays, but abandon the assumption of a unique vacuum element.

This note is organized as follows. Section 2 describes three of the axioms that are required for any Quantum Field Theory. We then show that the axioms are mutually inconsistent. Section 3 briefly discusses alternative subsets of these axioms which would be internally consistent. Section 4 concludes.

**2. The axioms and the main result**

An axiom of Quantum Field Theory is that the vacuum state is a vector in an infinite-dimensional Hilbert space.

**Assumption 1.** The vacuum, $\Psi$, is a vector in an infinite-dimensional Hilbert space $\mathcal{H}$.

Hilbert space is a separable Euclidean space which is complete and infinite-dimensional. It is well-known that every separable Euclidean space has a countable orthonormal basis (for example, p.148 of [1]). Any vector in Hilbert space can be written in terms of an infinite-dimensional orthogonal basis $\{\psi_\alpha\}$

$$\Psi = c_1\psi_1 + c_2\psi_2 + ...$$

where $c_1, c_2,...$ are scalar constants.

Quantum Field Theory requires the existence of a vacuum state which is everywhere translationally invariant. This is formalized in the usual sense as follows.

**Assumption 2.** The vacuum state is invariant to the inhomogenous Lorentz (Poincare) group) $SL(2,C)$. For any translation, $a$ and any rotation, $\Lambda$, we have

(1) $$T_{a,\Lambda}\Psi = \Psi.$$

We make no use of the Lorentz transformations in this paper. Only translational invariance is needed in order to derive a contradiction.

It follows immediately from assumptions 1 and 2 that the vacuum state must be constant in a particular sense.

**Proposition 1.** Assumptions 1 and 2 imply that the vacuum wavestate must be constant over the orthogonal basis,

$$\Psi = (c_1\psi_1 + c_1\psi_2 +,...)$$

for some constant $c_1$.

*Proof.* A pure translation leaves the state unchanged,

(2)
$$T_a(c_1\psi_1, c_2\psi_2,...) = (c_{1+a}\psi_1, c_{2+a}\psi_2,...)$$
$$= (c_1\psi_1, c_2\psi_2,...)$$

for all $a$. This implies that $c_{1+a} = c_1$ for all $a$. ∎

Assumptions 1 and 2 describe an infinite-dimensional vector which is everywhere translationally invariant. The invariance implies a wave state which is a constant over its basis. Unfortunately, this necessarily implies that the vacuum has a norm which is either divergent or identically zero.

**Proposition 2.** Assumptions 1 and 2 imply the vacuum state has a norm which is divergent or identically zero.

*Proof.* The norm of the vacuum $\|\Psi\|$ is given by

(3)
$$(\Psi, \Psi) = ((c_1\psi_1 + c_1\psi_2...), (c_1\psi_1 + c_1\psi_2...)).$$

A property of any Hilbert space norm is that

(4)
$$(x, y + z) = (x, y) + (x, z)$$

for any vectors $x, y, z \in \mathcal{H}$. Using (4), the norm in (3) becomes

(5)
$$c_1^2 + c_1^2 + ....$$

which is zero if $c_1 = 0$ and divergent otherwise. ∎

**Corollary.** No infinite-dimensional Hilbert space contains a vector which is everywhere translationally invariant and is also a unit ray.

This presents a problem for constructing a Quantum Field Theory in the sense of [2, 3]. All such formulations of QFT assume the existence of an invariant vacuum state with unit norm, a mathematical impossibility.

### 3. Alternative Formulations

On purely mathematical grounds, a Quantum Field Theory cannot be constructed from a unique vacuum element with norm one. In this section, we consider the consequences of relaxing the assumptions in order to construct a consistent theory.

**Formulation I.** There is an invariant element $\Psi_0$ but it does not have unit norm.

Suppose we were to relax the assumption of a unit norm but maintain the other assumptions. There is a vector in an infinite-dimensional Hilbert space which is invariant to pure translations. From proposition 2, the norm of this invariant state is zero or infinity. The vacuum state in this case would not have a finite norm.

In particular if $(\Psi_0, \Psi_0) = 0$, then the vacuum expectations

(6) $$\langle \Psi_0, \varphi_1 \varphi_2 \cdots \varphi_n \Psi_0 \rangle = 0$$

for any Hermitian field operators $\varphi_j$ defined on (a domain of) elements in $\mathcal{H}$. The expectations collapse into zeros because the states are constructed from a vacuum element which itself has norm zero. Similarly if $(\Psi_0, \Psi_0) = \infty$, the vacuum element has divergent norm, and the vacuum expectations all become divergent.

**Formulation II.** All states in the theory are unit rays in Hilbert space but there is no Poincare invariant vacuum state.

This case is consistent with classical Quantum Mechanics. The absence of a unique vacuum element would seem to preclude construction of a Fock-like space. It would thus be difficult to reconcile with any Quantum Field Theory. However, relativistic Quantum Mechanical calculations for fixed-particle number problems are not affected.

## 4. Conclusions

In a Quantum Field Theory, the quantum fields are operator valued functions which act in a Hilbert space $\mathcal{H}$ and obey certain axioms. The classical axioms of [3] require that $\mathcal{H}$ contain a vacuum vector which has unit norm and is invariant under the representation of the Poincare group. In this note, we show that this set of requirements is mathematically impossible in an infinite dimensional Hilbert space.

To derive a theory that is internally consistent, one of the axioms must be weakened. We briefly consider two possibilities. The first is to insist on a unique vacuum element, but with a length which is zero or infinity. The second is to insist all states have a norm of one, but abandon the requirement of a unique vacuum element. Of these two possibilities, the latter is far more convenient because it is consistent with Quantum Mechanics.

# References

[1] Kolmogorov, A. N. and S. V. Fomin, *Introductory Real Analysis*, (Dover Publications, New York, 1975)

[2] Osterwalder, K. and R. Schrader, "Axioms for Euclidean Green's functions", *Commun. Math. Phys.* **31**, (1973); **42**, 4, (1975)

[3] Streater, R. and Wightman, A., *PCT, SPIN and Statistics and all That*, (W.A. Benjamin, New York, 1964)

[4] Weinberg, S., *The Quantum Theory of Fields*, (Cambridge University Press, Cambridge, 2002).